\documentclass[twocolumn,epjc3]{svjour3}  

\usepackage{graphicx}
\usepackage{physics}
\usepackage{amssymb}
\usepackage{xcolor}

\journalname{Eur. Phys. J. A}

\newcommand{\unit}[1]{\;\mathrm{#1}}
\newcommand{\me}[0]{\mathrm{e}}
\newcommand{\iu}[0]{\mathrm{i}}

\renewcommand{\vec}[1]{\boldsymbol{#1}}
\usepackage{ulem}
\begin{document}

\title{On nuclear coalescence in small interacting systems}

\author{M. Kachelrie\ss \thanksref{addr1}
        \and
        S. Ostapchenko\thanksref{addr3}
        \and
        J. Tjemsland\thanksref{addr1}
}

\institute{Institutt for fysikk, NTNU, Trondheim, Norway
\label{addr1}
\and
 D.V. Skobeltsyn Institute of Nuclear Physics,
 Moscow State University, Moscow, Russia\label{addr3}
}

\date{Received: date / Accepted: date}

\maketitle

\begin{abstract}
The formation of light nuclei can be described as the coalescence of clusters
of nucleons into nuclei. In the case of small interacting systems, such as
dark matter and $e^+e^-$ annihilations or $pp$ collisions, the coalescence
condition is often imposed only in momentum space and hence the size of the
interaction region is neglected. On the other hand, in
most coalescence models used for heavy ion collisions, the coalescence
probability is controlled mainly by the size of the interaction region, while
two-nucleon momentum correlations are either neglected or
treated as collective flow.
Recent experimental data from $pp$ collisions at LHC have been interpreted as
evidence for such collective
behaviour, even in small interacting systems. We argue that these data are
naturally explained in the framework of conventional QCD inspired event
generators when both
two-nucleon momentum correlations and the size of the hadronic emission
volume are taken into account. To include both effects, we employ a per-event
coalescence model based on the Wigner function representation of the produced
nuclei states. This model reproduces well the source size for baryon emission
and the coalescence factor $B_2$ measured recently by the ALICE collaboration
in $pp$ collisions. 

\end{abstract}
\section{Introduction}
\label{sec:introduction}

The production mechanism for light clusters of nucleons, such as deuteron,
helium-3, tritium and their antiparticles, in particle interactions has
recently attracted increased attention from both the astroparticle and
heavy ion communities. In heavy ion collisions, their small binding energies
make these particles sensitive probes for two-nucleon correlations and density
fluctuations, which may shed light on the QCD phase
diagram~\cite{Caines:2017vvs}.
These particles are also of particular 
interest in cosmic ray studies, because the expected low astrophysical
backgrounds makes them ideal probes for exotic physics~\cite{Donato:1999gy}.
Furthermore, the sensitivities of the AMS-02 and GAPS
experiments~\cite{battiston_antimatter_2008,Aramaki:2015laa}  are close to
the expected fluxes of antideuterons from secondary production and, for
optimistic parameters, from dark matter
annihilations~\cite{vonDoetinchem:2020vbj}. In order to correctly interpret 
the results of these experiments, a precise description of the production mechanism
of light nuclei\footnote{%
Most of the discussions in this work apply equally well to particles
as to antiparticles, and the prefix ``anti'' will thus often be dropped.}
 is important. 

In small interacting systems, such as dark matter and $e^+e^-$  annihilations
or $pp$ collisions, the production of light nuclei is usually described by
the coalescence model in momentum
space~\cite{Schwarzschild:1963zz,butler_deuterons_1963,Chardonnet:1997dv}, 
where nucleons originating from a particle collision
merge to form a nucleus if their invariant momentum difference is
smaller than the coalescence momentum $p_0$. Traditionally, the 
yield of a nucleus with mass number $A=Z+N$ and charge $Z$ has been linked
to the yields of protons $p$ and neutrons $n$ via the coalescence factor
$B_A$ as
\begin{equation}
    E_A\dv[3]{N_A}{P_A} = B_A\left(E_p\dv[3]{N_p}{p_p}\right)^Z
    \left(E_n\dv[3]{N_n}{p_n}\right)^N.
\label{eq:BA}
\end{equation}
Here, $P_A/A= p_n= p_p$ is the momentum of the nucleus and nucleons,
respectively. In the limit of isotropic nucleon yields, the relation
between $B_A$ and $p_0$ is
\begin{equation}
    B_A = A\left(\frac{4\pi}{3}\frac{p_0^3}{m_p}\right)^{A-1}.
    \label{eq:BA_isotropic}
\end{equation}
This picture can be improved by taking into account two-particle 
correlations provided by Monte Carlo event generators for strong
interactions, as proposed
in Refs.~\cite{Dal_thesis,kadastik_enhanced_2010}. Such two-particle
correlations are especially important in small interacting systems,
since there the nucleon yields deviate strongly from isotropy. This
approach is commonly used to predict the antinucleus yield in cosmic ray
interactions, as well as from dark matter decays or
annihilations~\cite{Cui:2010ud,ibarra_prospects_2013,%
Fornengo:2013osa,Dal:2014nda,delahaye_antideuterons_2015,Herms:2016vop,%
Coogan:2017pwt,Lin:2018avl,Carlson:2014ssa,Cirelli:2014qia,Shukla:2020bql,%
Li:2018dxj}, for a recent review see
Ref.~\cite{vonDoetinchem:2020vbj}. 
In order to be predictive, $B_A$ and $p_0$ must be independent of the
centre-of-mass (c.m.) energy and the interaction process. However, the latter
is not the case if the coalescence condition is only imposed in momentum
space, since then the process dependent size of the formation region is
neglected.

An alternative scheme where the coalescence condition is imposed in position
space is often employed for heavy ion
collisions~\cite{Csernai:1986qf,Nagle:1996vp}.
Here, the coalescence factor scales with the volume of the emission region
of hadrons as $B_A\propto V^{A-1}$. Much efforts have been spent on
unifying these pictures using, e.g., Wigner functions~\cite{scheibl_coalescence_1999} and imposing the coalescence condition in 
phase space, see Ref.~\cite{Danielewicz:1991dh} for a review
of early works. Such models differ mainly in the way the Wigner
  function of the nucleons is determined: The phase-space distributions of
  nucleons used in the coalescence models may be obtained, e.g., from transport
  models like the AMPT scheme~\cite{Sun:2020uoj} or hybrid schemes combining
  a hydrodynamical with a microscopic hadron cascade model~\cite{Zhao:2020irc}.
  Alternatively, analytical coalescence formula like the COAL-SH
  scheme~\cite{Sun:2017xrx} or statistical models which relate the phase-space
  volume at kinetic freeze-out to the entropy per nucleon have been
  proposed~\cite{Sun:2018jhg}.
  Finally, Refs.~\cite{Shuryak:2018lgd,Shuryak:2019ikv}, have 
  studied the influence of preclustering of baryons due to nucleon
  interactions on the coalescence process. 

A key observation in all approaches relying on the phase space
picture is that the coalescence probability
depends on the size of the hadronic emission region,
which can be measured in femtoscopy (often also called Hanburry-Twiss-Brown)
correlation experiments~\cite{scheibl_coalescence_1999}.
This connection has recently been applied to $pp$ collisions, both in cosmic ray
~\cite{Blum:2017qnn} and LHC studies~\cite{Blum:2019suo,Bellini:2020cbj}.
In particular, it was argued in Ref.~\cite{Bellini:2020cbj} that the success
of the femtoscopy analysis is  strong evidence that 
coalescence is the major
production mechanism of light nuclei. Moreover, these authors suggested
that the use of experimental data from femtoscopy correlation experiments
allows one to reliably predict the yield of light antinuclei in cosmic ray
interactions, thereby avoiding the need of additional theoretical inputs.

The approaches discussed above are all based on the coalescence picture,
but differ on how the coalescence condition is implemented
and how the two-nucleon states are determined. 
In a competing approach one employs statistical thermal
models~\cite{Acharya:2017bso,Andronic:2017pug,Vovchenko:2018fiy,Bellini:2018epz,Chen:2018tnh,Xu:2018jff,Oliinychenko:2018ugs}.
Here one assumes that both  hadronisation and the formation of light nuclei
occurs as a chemical freeze-out process in a radially expanding ``fireball''
of a Quark-Gluon Plasma (QGP). These models are motivated by the observation
that the spectra of light nuclei 
are consistent with a thermal distribution, with the same freeze-out temperature
as for mesons and nucleons~\cite{Andronic:2017pug}. 
Intriguingly, experimental data from collisions of small systems, such
as $pp$ and $p$Pb, show features characteristic for collective flows, or even for the
formation of a QGP, see Ref.~\cite{Nagle:2018nvi} for a review. 
It has therefore been suggested that the thermal production of
light nuclei can be applied even to small interacting
systems~\cite{vonDoetinchem:2020vbj,Bellini:2018epz,Cleymans:2011pe,Acharya:2020sfy}. However, it is difficult
to reconcile how the nuclei with their small binding energies survive the
chemical freeze-out. Even more, the energy spectrum of the nucleons
is in the coalescence picture inherited by the nuclei (up to a quantum 
mechanical correction factor~\cite{Csernai:1986qf}), and the apparent
quasi-thermal spectra of light nuclei can therefore be explained by
coalescence as well.

In Refs.~\cite{Kachelriess:2019taq,Kachelriess:2020uoh}, we developed
a coalescence model based on the Wigner function representation of the 
produced nuclei states, which includes two-nucleon momentum
correlations obtained from QCD inspired event generators
(we will use the abbreviation WiFunC,
i.e.\  Wigner Functions with Correlations, 
for this model). In this work, we argue that neither two-particle correlations
nor the source size can be neglected when describing the cluster formation
in small interacting
systems\footnote{For concreteness, we will only discuss
the production of deuterons, but the same considerations also hold for 
larger clusters of nucleons with small binding energies,
such as helium-3 and tritium.}.
Furthermore, we will use this model to describe the production of hadrons
and nuclei in high energy $pp$ collisions and compare it to recent
experimental data by the ALICE collaboration on the size of the
baryon emitting source~\cite{Acharya:2020dfb} and on the multiplicity and
transverse momentum dependence of the coalescence factor
$B_2$~\cite{collaboration_production_2018,Acharya:2020sfy,Acharya:2019rgc}.
Both data sets have been interpreted as evidence of collective flows, but
we will show that the same characteristics are described using QCD
inspired event generators, like QGSJET~II~\cite{Ostapchenko:2010vb,Ostapchenko:2013pia} and Pythia~8.2~\cite{sjostrand_pythia_2006,sjostrand_introduction_2015}.
Finally, we comment on the suggestion that femtoscopy data alone are sufficient
to predict the yield of light antinuclei for astrophysical applications.

This paper treats several different topics related to the formation of 
nuclei by the coalescence mechanism
in small interacting systems, with a focus on recent experimental
data, and is structured as follows. We review the WiFunC model in
Section~\ref{sec:qm} and its relation to the femtoscopy framework in
Section~\ref{sec:femtoscopy}. In Section~\ref{sec:source}, we
compare our predictions for the size of the baryon emitting source 
to recent measurements of the ALICE collaboration in a femtoscopy experiment.
In Section~\ref{sec:multiplicity}, the multiplicity and transverse momentum
dependencies of the coalescence factor $B_2$ in $pp$ collisions at  13\,TeV,
measured by the ALICE collaboration, are compared to the WiFunC model. In
Section~\ref{sec:astrophysical_applications}
we make comments on the use of isotropic models in astrophysical applications.

\section{The quantum mechanics of coalescence and the WiFunC model}
\label{sec:qm}

The WiFunC model is based on the quantum mechanical description of the coalescence
process reviewed in, e.g., Refs.~\cite{scheibl_coalescence_1999,Bellini:2018epz}.
Here we will highlight only the main steps. In this approach, the final state produced in a
particle collision is described by a density matrix. Thus, one can find
the deuteron spectrum in the sudden approximation by
projecting the deuteron density matrix, $\rho_d=\ket{\phi_d}\bra{\phi_d}$, onto
the reduced density matrix
$\rho_\mathrm{nucl}=\ket{\psi_p\psi_n}\bra{\psi_p\psi_n}$ describing the
coalecsing nucleons,
\begin{equation}
  \dv[3]{N_d}{P_d}=\tr{\rho_d\rho_\mathrm{nucl}}.
\end{equation}
By factoring out the  c.m.\ motion of the deuteron, 
$\phi_d\propto\exp{\iu \vec P_d\cdot \vec{r}_d}\varphi_d$, 
one can show that
\begin{equation}
\begin{aligned}
    \dv[3]{N_d}{P_d} = \frac{3}{8}
    \int&\frac{\dd[3]{r_d}\dd[3]{r}\dd[3]{q}}{(2\pi)^6}
    \mathcal{D}(\vec{r}, \vec{q})\\
    &\times W_{np}(\vec{P}_d/2+
    \vec{q}, \vec{P}_d/2-\vec{q}, \vec{r}_n, \vec{r}_p),
    \label{eq:main_integral}
\end{aligned}
\end{equation}
where the statistical factor $3/8$ arises from averaging over spin and
isospin and $\vec{r}\equiv\vec{r}_n-\vec{r}_p$. Here, 
\begin{equation}
    \mathcal{D}(\vec{r}, \vec{q})=\int\dd[3]{\xi}\me^{-\iu\vec{q}\cdot\vec \xi}
    \varphi_d(\vec{r}+\vec{\xi}/2)\varphi_d^*(\vec{r}-\vec{\xi}/2)
\end{equation}
is the deuteron Wigner function, $W_{np}$ is the Wigner function of the
two-nucleon state, and $\varphi_d$ is the internal deuteron wave function. 
If one approximate the deuteron wave function as a Gaussian, then
$\mathcal{D}(\vec{r}, \vec{q})=8\exp{-r^2/d-q^2d^2}$, with
$d\simeq 3.2\unit{fm}$. However, apart from analytical estimates a more
accurate wave function should be used, such as a two-Gaussian fit
to the Hulthen wave function, chosen in Ref.~\cite{Kachelriess:2019taq}.

To proceed, one has to specify the Wigner function $W_{np}$ 
in Eq.~\eqref{eq:main_integral}. One possibility is to use simulations
in order to determine the phase-space distribution of nucleons. Both the
perturbative and non-perturbative evolution in Monte Carlo generators
of strong interactions are based on momentum eigenstates and, hence, they
provide only information on momentum correlations of nucleons.
The addition of spatial information requires thus the transition to a
semi-classical picture. Alternatively, one can neglect two-nucleon correlations 
and assume an isotropic source, as it is often done when describing heavy ion
collisions. Finally, one can derive two-particle correlations from
experimental data. This is the approach used in the femtoscopy framework
that will be discussed in the next section.

The first case is used in the WiFunC model~\cite{Kachelriess:2019taq}
which combines two-nucleon momentum
   correlations obtained from QCD inspired event generators, with 
 a simple analytical model for the spatial distribution of nucleons.
 Assuming
a factorisation of the momentum and position dependence in the Wigner function,
\begin{equation}
W_{np}=
H_{np}(\vec{r}_n, \vec{r}_p)G_{np}(\vec{P}_d/2+\vec{q}, \vec{P}_d/2-\vec{q}),
\label{eq:ansatz_1}
\end{equation}
as well as neglecting spatial correlations,
$H_{np}(\vec r_n, \vec r_p)=h(\vec r_n)h(\vec r_p)$,
and choosing a Gaussian ansatz for the spatial distribution,
\begin{equation}
    h(\vec r) = \left(2\pi \sigma^2\right)^{-3/2}\exp{-\frac{r^2}{2\sigma^2}},
    \label{eq:ansatz_2}
\end{equation}
Eq.~\eqref{eq:main_integral} becomes
\begin{equation}
     \dv[3]{N_d}{P_d} = \frac{3\zeta}{(2\pi)^6}\int \dd[3]{q}\me^{-q^2d^2}
     G_{np}(\vec{P}_d/2+\vec{q}, \vec{P}_d/2-\vec{q}).
     \label{eq:WiFunC}
\end{equation}
The function $\zeta$ reflects the spatial distribution of the nucleons,
and is thus clearly process dependent. It is in general given by
\begin{equation}   
    \zeta(\sigma_\parallel, \sigma_\perp, d) =
    \sqrt{\frac{d^2}{d^2+4\tilde{\sigma}_\perp^2}}
    \sqrt{\frac{d^2}{d^2+4\sigma_\perp^2}}
    \sqrt{\frac{d^2}{d^2+4\sigma_\parallel^2}},
    \label{eq:zeta}
\end{equation}
where $\tilde{\sigma}_\perp^2=
\sigma_\perp^2/(\cos^2\theta + \gamma^2\sin^2\theta)$.
Here we distinguished between the longitudinal and transverse spreads
$\sigma_{\|,\perp}$ of the emission volume. The transverse spread is modified
when boosting from the c.m.\ frame of the original
particle collision to the deuteron frame. Thus $\gamma$ is the Lorentz factor
of the produced deuteron in the collider frame,  while $\theta$
is the angle between the deuteron momentum and the beam axis.
Note that, in contrast to our earlier treatement in
Refs.~\cite{Kachelriess:2019taq,Kachelriess:2020uoh},
we have  included the Lorentz boost in only one of the two
transverse components: If the $xy$ cordinates are rotated such that
$\vec P_d$ is contained in, e.g., the $xz$ plane, then the $\sigma_y$
component will not be affected by the Lorentz boost.

Nucleon momentum correlations are provided by the event generator, 
while the process dependence is incorporated in the spread $\sigma$. The
spread will in general have a geometrical contribution due to a finite
spatial extension of
the colliding particles, and a contribution related to the perturbative
cascade and hadronisation,
\begin{equation}
    \sigma_{\parallel,\perp}^2 = \sigma_{\parallel,\perp(e^{\pm})}^2
    +\sigma_{\parallel,\perp(\mathrm{geom})}^2.
\end{equation}
The geometrical contributions can be approximated as
\begin{equation}
\begin{aligned}
    &\sigma_{\perp(\mathrm{geom})}^2\simeq \frac{2R_1^2R_2^2}{R_1^2 + R_2^2},\\
    &\sigma_{\parallel(\mathrm{geom})}^2\simeq \max\{R_1, R_2\},
    \label{eq:interaction_process}
\end{aligned}
\end{equation}
where $R_1$ and $R_2$ are the radii of the colliding particles, while the
point-like contributions are given by
$\sigma_{\parallel(e^{\pm})}\simeq R_p\simeq 1\unit{fm}$ and
$\sigma_{\perp(e^{\pm})}\simeq \Lambda_\mathrm{QCD}^{-1}\simeq 1\unit{fm}$.
This simple picture is expected to give accurate results for $pp$ 
interactions,
while in the case of $pA$ and $AA$ collisions the geometrical contribution
varies from event to event: While peripheral interactions which are dominated
by binary collisions between a pair of projectile and target nucleons are
characterised by $\sigma_{\|(\mathrm{geom})}\simeq R_p$, the size may increase
to $\sigma_{\|(\mathrm{geom})}\simeq  R_A$ for  the most
central collisions. Consequently, the multiplicity of secondaries
and the size of the source region are strongly correlated.

Neglecting for the moment this correlation, and approximating 
the radius of a nucleus by
\begin{equation}
    R_A\simeq a_0A^{1/3},
    \label{eq:nucleus_radius}
\end{equation}
with $a_0\simeq 1.1\unit{fm}$, allows us to use only one free 
parameter, 
\begin{equation}
    \sigma\equiv \sigma_{(e^\pm)} = a_0 = \sigma_{(pp)}/\sqrt{2}\simeq 1\unit{fm},
\end{equation}
whose physical interpretation is the size of the
emission region of nucleons.

Ideally, also the position integral in Eq.~\eqref{eq:main_integral}
should be evaluated event-by-event. It is therefore worth pointing out
that some event generators like Pythia~8.2 have implemented semi-classical
trajectories of the produced hadrons~\cite{Ferreres-Sole:2018vgo}. Thus,
using Pythia one can instead directly evaluate
\begin{equation}
    \dv[3]{N_d}{P_d} = 3
    \int\frac{\dd[3]{r}\dd[3]{q}}{(2\pi)^6}
    \me^{-r^2/d^2 - q^2d^2}
    W_{np}(\vec{p}_p, \vec{p}_n, \vec{r}_p, \vec{r}_n).
    \label{eq:pythia_model}
\end{equation}
relying on the semi-classical description of the spatial correlations
provided by the simulation. 
A simple model applying a hard cut-off in both
momentum and position space has been considered using UrQMD in
Ref.~\cite{Sombun:2018yqh}.
The approach of the WiFunC model could be carried over in straight-forward way 
to these models, replacing the hard cutoffs with Eq.~(\ref{eq:pythia_model}).

Because of the generality of Eq.~\eqref{eq:WiFunC}, the WiFunC model can in
principle be used to describe the production of other nucleus-like systems
with small binding energies if the approximate wave function of the
produced system is known. One additional application of the WiFunC model could
therefore be the production of exotic bound states such as the $X(3872)$
or the $Z_{cs}(3985)$, if they are deuteron-like bound states~\cite{Esposito:2014rxa,Guo:2017jvc,Kalashnikova:2018vkv,Yamaguchi:2019vea,Sun:2020hjw,Liu:2020nge}.

\section{Relation to the femtoscopy framework}
\label{sec:femtoscopy}

The emission volume probed in femtoscopy correlation experiments 
is directly linked to the  distribution of nucleons, and
can thus be used to check the validity of the WiFunC model. 
In a similar fashion, the emission volume can be related to the
coalescence factor $B_A$, as was done in
Refs.~\cite{scheibl_coalescence_1999,Blum:2019suo,Bellini:2020cbj}. 
However, in order to derive their analytic relationship, the so-called
smoothness approximation~\cite{Lisa:2005dd} was applied on top of the sudden
approximation used in the previous section. In this approximation,
the $q$ dependence in the nucleon Wigner
function is assumed to be negligible so that the $q$ integral in
Eq.~\eqref{eq:main_integral} can be evaluated. As remarked in
Ref.~\cite{Bellini:2020cbj}, this may be justified for heavy ion collisions
where the size of the produced nuclear clusters can be neglected compared
to the size of the emitting source. However,
a more careful treatment is warranted for small interacting systems.
To see this, we note that  applying  the sudden approximation to
Eq.~\eqref{eq:WiFunC} implies that two-nucleon correlations are  neglected,
but these correlations should be kept for small interacting systems. The WiFunC
model evades these problems because it evaluates the momentum integral
using the momentum distributions supplied by an event generator.

Within the smoothness approximation, the deuteron spectrum 
\eqref{eq:main_integral} can be written as
\begin{equation}
\begin{aligned}
    \dv[3]{N_d}{P_d} = &\frac{3}{8}
    \int\frac{\dd[3]{r}}{(2\pi)^3}
    |\varphi_d(\vec r)|^2\\
    &\times\int \dd[3]{r_d} 
    W_{np}(\vec{P}_d/2, \vec{P}_d/2, \vec{r}_n, \vec{r}_p),
\end{aligned}
\end{equation}
while the nucleon spectra are given by%
\footnote{Notice that we have included here, in contrast to the coalescence
  factor~\eqref{eq:BA},  two-nucleon correlations.
  Since typically only the proton spectra will be available experimentally,
  it is common to assume factorised nucleon distributions. As the correlations
  are provided by Monte Carlo simulations and are included in the WiFunC model,
  we keep them in this expression.}
\begin{equation}
\begin{aligned}
    \frac{\dd[6]{N}}{\dd p_p^3\;\dd p_n^3}=&\int\frac{\dd[3]r}{(2\pi)^6}\int\dd[3]r_d
    W_{np}(\vec p_p, \vec p_n, \vec r_p, \vec r_n).
\end{aligned}
\end{equation}
Following the authors of Refs.~\cite{Blum:2019suo,Bellini:2020cbj}, we 
assume for simplicity $E_d/(E_pE_n)=2/m_N$ in the deutron rest frame. 
Then the coalescence factor~\eqref{eq:BA} becomes
\begin{equation}
  B_2(\vec P_d) \simeq \frac{3(2\pi)^3}{2m} \int \dd[3]{r}|\varphi_d(\vec r)|^2
    \mathcal{S}_2(\vec r, \vec P_d),
    \label{eq:B2_blum}
\end{equation}
with the source function defined as
\begin{equation}
    \mathcal{S}_2(\vec r, \vec P_d) = 
    \frac{\int\dd[3]r_dW_{np}(\vec P_d/2, \vec P_d/2, \vec r_p, \vec r_n)}{
    \int\dd[3]r_d\dd[3]r W_{np}(\vec P_d/2, \vec P_d/2, \vec r_p, \vec r_n)
    }.
    \label{eq:C_wifi}
\end{equation}
Measured particles will always be affected by final state interactions. This
significantly affects two-particle correlation experiments: Even from initially
uncorrelated particles one will measure a correlation
\begin{equation}
    \mathcal{C}(\vec q) = \int \dd[3]{r} S(\vec r) |\Psi(\vec r, \vec q)|^2,
    \label{eq:C_exp}
\end{equation}
where $S(\vec r)$ is the emission source function and the final state
interactions are encoded in the wave function $\Psi$~\cite{Lisa:2005dd}.
This is very similar to Eqs.~\eqref{eq:WiFunC} and~\eqref{eq:B2_blum}:
Coalescence is effectively a final 
state interaction that affects the two-nucleon correlations. 

The authors of Refs.~\cite{Blum:2019suo,Bellini:2020cbj} used
Eq.~\eqref{eq:B2_blum} to derive numerical estimates of the $B_2$ factor
as a function of the source radius $r$ measured in femtoscopy experiments.
This approach looks very promising, since it allows one to express the
coalescence factor only in terms of measurable quantities.
Unfortunately, any numerical evaluation is additionally based on three
assumptions on the two-nucleon wave function: i) the spatial distribution
has to be prescribed,  ii) its characteristic size is assumed to be much
larger than the one of the produced antinucleus states, such that the
smoothness approximation can be used, iii) the two-nucleon
momentum correlations are negligible. Yet, all these assumptions are generally
not valid for collisions of small systems,  as correctly noted already in
Ref.~\cite{Blum:2019suo}.
Furthermore, the correlation function has to be inferred from experimental data,
and is thus only available for the central rapidity region.
The approximations required in the approach of
Refs.~\cite{Blum:2019suo,Bellini:2020cbj} 
are avoided in the WiFunC model, since the used Monte Carlo generators
provide two-nucleon momentum correlations which in turn leads to a
non-trivial source function.

\section{Size of baryon-emitting source}
\label{sec:source}

The source radius of the baryon emission  in $pp$
collisions at $13\unit{TeV}$ was recently measured by the ALICE collaboration,
assuming a Gaussian source profile,
\begin{equation}
    \mathcal{S}(|\vec r_p-\vec r_n|) \propto \exp{-\frac{(\vec r_p-\vec r_n)^2}{4r_0^2}},
    \label{eq:gaussian_source}
\end{equation}
using the femtoscopy framework, cf.\ with Eq.~\eqref{eq:C_exp} of
Ref.~\cite{Acharya:2020dfb}. Here, the distance
  $\vec r = \vec r_p-\vec r_n$ between the two nucleons is defined in their
  pair rest frame.
This study indicates that protons, antiprotons,
$\Lambda$ and $\bar\Lambda$ originate from the same source volume. Furthermore,
a decrease in the source size with increasing
transverse mass was observed. This decrease is often attributed to a collective
flow, but is, as we will see next, also naturally described in the WiFunC model.

Inserting the Gaussian ansatz for the spatial distribution of
nucleons~\eqref{eq:ansatz_2} 
into the expression~\eqref{eq:C_wifi} for the
source leads to
\begin{equation}
    \mathcal{S}_2(r)\propto \int \dd[2]\Omega
    \exp{-\frac{r_z^2}{4\sigma_\parallel^2}
         -\frac{r_y^2}{4\sigma_\perp^2}
         -\frac{r_x^2}{4\sigma_\perp^2}\frac{m_T^2}{m^2}} ,
    \label{eq:source_init}
\end{equation}
where we have taken into account that the Wigner functions and
  their spread, cf.\ with Eq.~(\ref{eq:zeta}), are defined in the collider
  frame.
Moreover, we chose the coordinate system such that $\hat{z}$ is directed along
the
initial beam direction and $\hat{y}$ is perpendicular to both $\hat{z}$ and
$\vec{P}_d.$ Furthermore, we used the identity
$m_T^2/m^2=\gamma^2\sin^2\theta+\cos^2\theta$, $m_T$ being the transverse mass.
Using the polar coordinates $r_x/r=\sin\varphi\sin\vartheta$ and
$r_y/r=\cos\varphi\sin\vartheta$, we find
\begin{equation}
    \mathcal{S}_2(r) \propto \me^{-r^2/4\sigma_{\parallel}^2}\times
    \mathcal{I}(r,m_T, \sigma_\parallel, \sigma_\perp),
    \label{eq:source_final}
\end{equation}
with
\begin{equation}
  \mathcal{I}(r,m_T, \sigma_\parallel, \sigma_\perp)=
    \int_0^{2\pi}\!\!\!\!\dd{\phi}\int_0^\pi\!\!\!\dd{\vartheta}\sin\vartheta
    \exp\left(-\frac{r^2\sin^2\vartheta}{4\sigma_\parallel^2} \mathcal{F} \right)
\end{equation}
and
\begin{equation}
  \mathcal{F} =  \cos^2\varphi\left(
    \frac{\sigma_\parallel^2}{\sigma_\perp^2}-1\right) +
    \sin^2\varphi\left(\frac{\sigma_\parallel^2}{\sigma_\perp^2}
    \frac{m_T^2}{m^2}-1\right) .
\end{equation}
Hence the WiFunC model predicts a non-trivial source function described
by a Gaussian source modified by the function
$\mathcal{I}(r,m_T, \sigma_\parallel, \sigma_\perp)$.

In order to compare the predicted source function to the measurement by the
ALICE collaboration, Eq.~\eqref{eq:source_final} must be compared to the
Gaussian source profile~\eqref{eq:gaussian_source} to fix $r_0(m_T)$.
In order to determine  $r_0(m_T)$, we perform a least-squares fit 
using as uncertainty $\mu\propto 1/\sqrt{\mathcal{S}_2(r)}$ as
the expected Gaussian error.  Additionally, we consider also a simple
analytical approximation:
By comparing the Taylor expansion of Eqs.~\eqref{eq:gaussian_source} 
and \eqref{eq:source_final}, one finds
\begin{equation}
    r_0^2/\sigma_\parallel^2 = 3\left[1+
    \left(\frac{m_T^2}{m^2} + 1\right)\frac{\sigma_\parallel^2}{\sigma_\perp^2}
    \right]^{-1} + \mathcal{O}(r^2/\sigma_\parallel^2).
    \label{eq:source_analytical}
\end{equation}

\begin{figure}[t]
    \centering
    \includegraphics[width=\columnwidth]{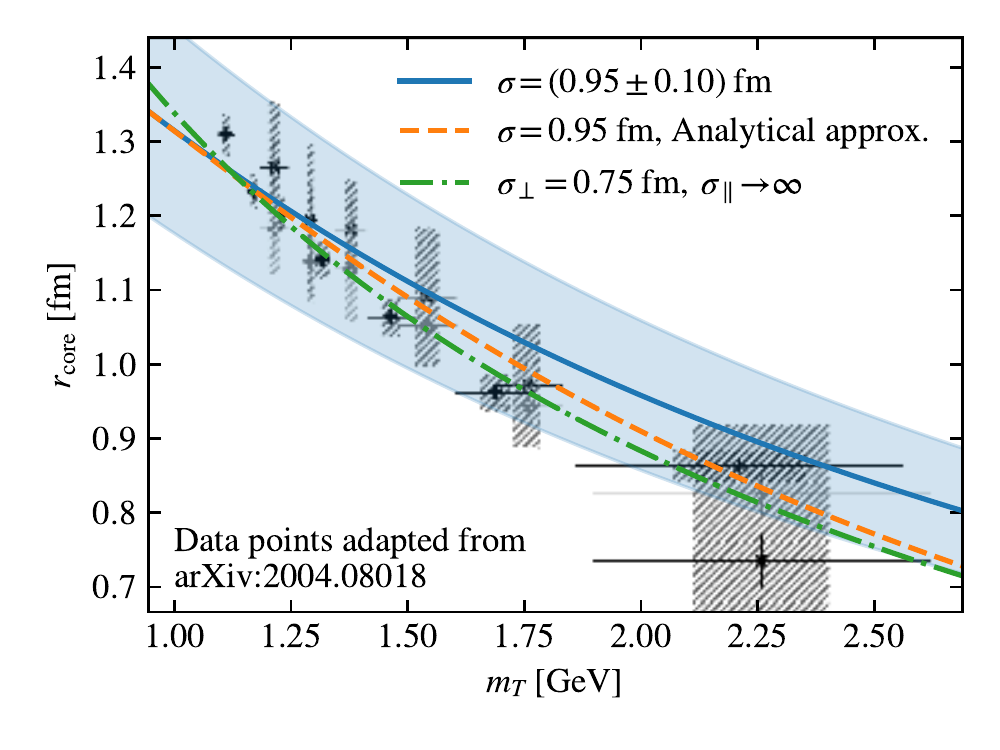}
    \caption{The Gaussian emission size predicted by the WiFunC model is
    compared to experimental data. The blue solid line shows the prediction
    of the WiFunC model; the shaded area corresponds to the uncertainty
    $\Delta\sigma=0.1\unit{fm}$. The simple analytical approximation
    in Eq. \eqref{eq:source_analytical} is shown as a dashed orange line.
    Finally, the green dashed dotted line indicates the limit
    $\sigma_\parallel\gg\sigma_\perp$.
    \label{fig:r_core}}
\end{figure}

In the analysis of the data on the source function in $pp$ collisions at
13\,TeV by  ALICE only high multiplicity events  (0--0.17\% $\mathrm{INEL}>0$)
were included~\cite{Acharya:2020dfb}. However,
the WiFunC model says that there is no (or only a weak) multiplicity dependence 
of the emission volume in $pp$ collisions.
In Fig.~\ref{fig:r_core}, we compare the source size $r_0$ estimated
for proton-proton pairs\footnote{A similar analysis can be done for
$\Lambda$ by changing $m_p\to m_\Lambda$. In this case, a correspondingly
larger $\sigma$ is expected.}, using both the exact source
function~\eqref{eq:source_final} (blue solid line) and the
approximation~\eqref{eq:source_analytical} (orange dashed line).
Additionally,  we show the source size obtained in the limit
$\sigma_\parallel\gg\sigma_\perp$ (green dashed-dotted line), which corresponds
to the steepest slope $r_0(m_T)$ possible in our model.
It is worth noticing that the data tend to give better fits
for $\sigma_\parallel>\sigma_\perp$, as expected from their physical
interpretations. Even so, we find not yet any need to fit them separately due
to the relatively large experimental uncertainties.

From Fig.~\ref{fig:r_core} one can infer $\sigma=(0.95\pm 0.1)$\,fm.
Intriguingly, the WiFunC model thus
describes the data well with values of $\sigma$ similar to those obtained in
Refs.~\cite{Kachelriess:2019taq,Kachelriess:2020uoh} by  a fit to
antideuteron measurements. More importantly, we have shown that the 
decrease of the source size with increasing transverse momentum, which is
often attributed to collective flows, is correctly described 
by the WiFunC model using QCD inspired MC generators.

\section{Multiplicity dependence of coalescence in small interacting systems}
\label{sec:multiplicity}

In the previous section, we focused on how the emission region of nucleons
is related to the source size measured in femtoscopy experiments.
Now we consider the effect of two-particle correlations on the
deuteron yield. To this end, we investigate how the coalescence factor $B_2$
of antideuterons measured at mid-rapidity ($|y|<0.5$) in $pp$ collisions at
13\,TeV depends on multiplicity and transverse momentum\footnote{We constrain
  this discussion to the data obtained at 13\,TeV because of their small
    experimental uncertainties, but the same qualitative features are seen
    also at 7\,TeV~\cite{Acharya:2019rgc}.}.

The experimental results are reported for a specific event class
($\mathrm{INEL}>0$) and are divided into different multiplicity classes in
terms of the percentage of the inclusive cross section, see
Ref.~\cite{Acharya:2020sfy} and references therein for their definition.
We aim to reproduce the data, generating inelastic $pp$ collisions at 13\,TeV
with QGSJET~II and Pythia~8.2, while  describing the coalescence
by the WiFunC model with $\sigma=0.9$\,fm, using
 the two-Gaussian wave function for the deutron.
We check the trigger condition
and classify the multiplicity class on an event-by-event basis. 
For comparison, 
we consider the standard per-event coalescence model with a hard
cutoff $p_0\sim 0.2\unit{GeV}$. This serves as a benchmark on what effects are
caused by particle correlations, and what by the source size in the WiFunC
model. 

\begin{figure}[htbp]
    \centering
    \includegraphics[width=\columnwidth]{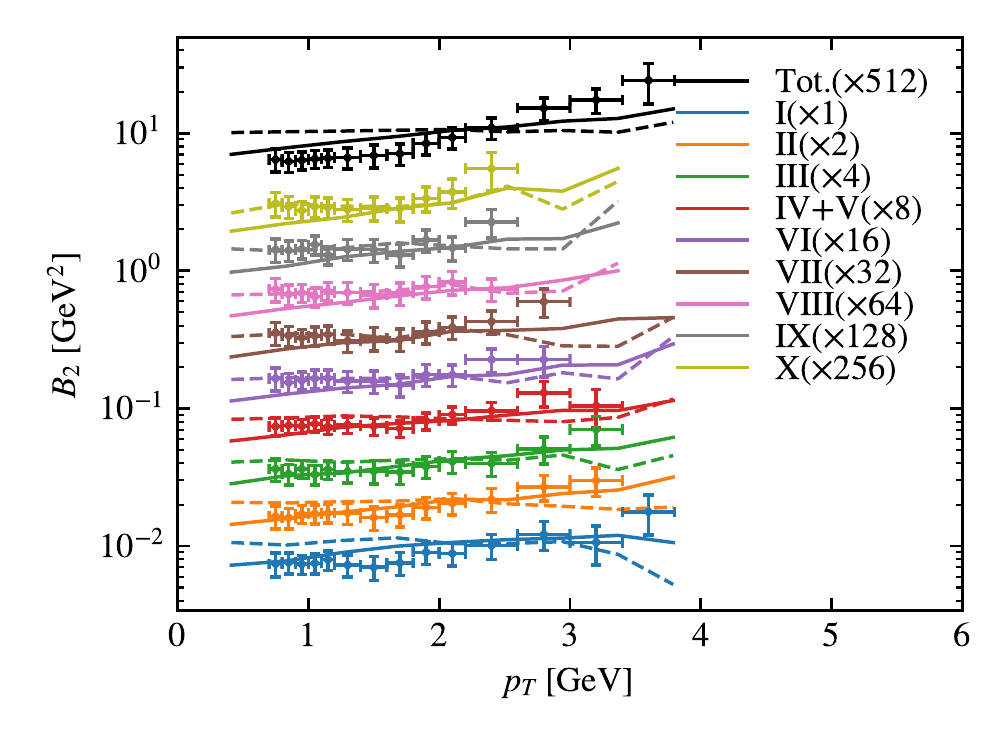}
    \vspace{-.8cm}\\
    \includegraphics[width=\columnwidth]{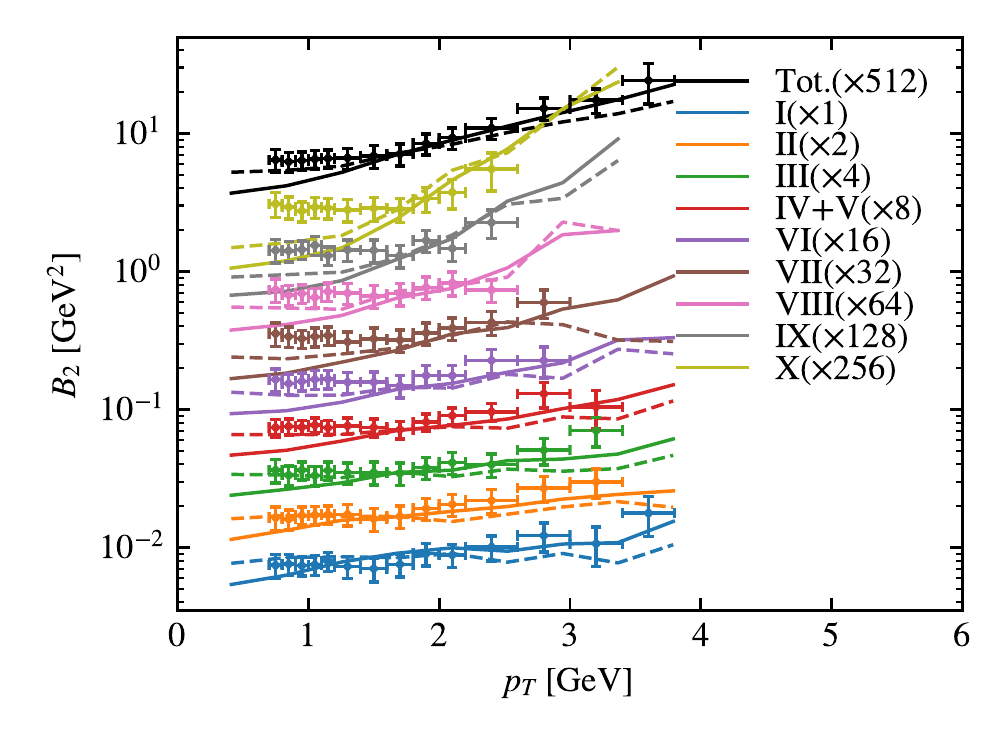}
    \caption{The coalescence factor $B_2$ for different multiplicity classes
    measured by the ALICE collaboration is compared to the predictions
    by QGSJET II (above) and Pythia 8.2 (below) using the WiFunC model (solid
    lines). The results for the standard coalescence model (dashed lines) are
    shown for comparison. Class I corresponds to largest multiplicities,
    while the multiplicity decreases with increasing class.}
    \label{fig:B2}
\end{figure}

The results are compared to the experimental data in Fig.~\ref{fig:B2}. Both
QGSJET II and Pythia 8 reproduce well the overall yield in the various
multiplicity classes. Furthermore, the qualitative behaviour of an increasing
transverse momentum $p_{T}$
slope of $B_2$ with increasing multiplicity is also reproduced. This increase
is often attributed to a collective flow, but our results indicate that it is
also well described by the WiFunC model combined with QCD inspired event
generators. While the overall
behaviour and trends of the experimental data are reasonably well
reproduced, deviations are expected as the event generators are not tuned to
two-particle correlations. Comparing the results from the WiFunC model, shown as
solid lines, to those of the standard coalescence model (dashed lines), one can
notice that the  multiplicity dependence of the slope of $B_2$  is stronger
in the WiFunC model. Even so, there is also an increase in the slope of $B_2$ in
the standard coalescence model, which is stronger in the case of Pythia.
This indicates that two-particle correlations, although not the only effect
responsible for the growing slope of $B_2$, are not negligible for 
$pp$ collisions in the kinematical range considered.

\begin{figure}[htbp]
    \centering
    \includegraphics[width=\columnwidth]{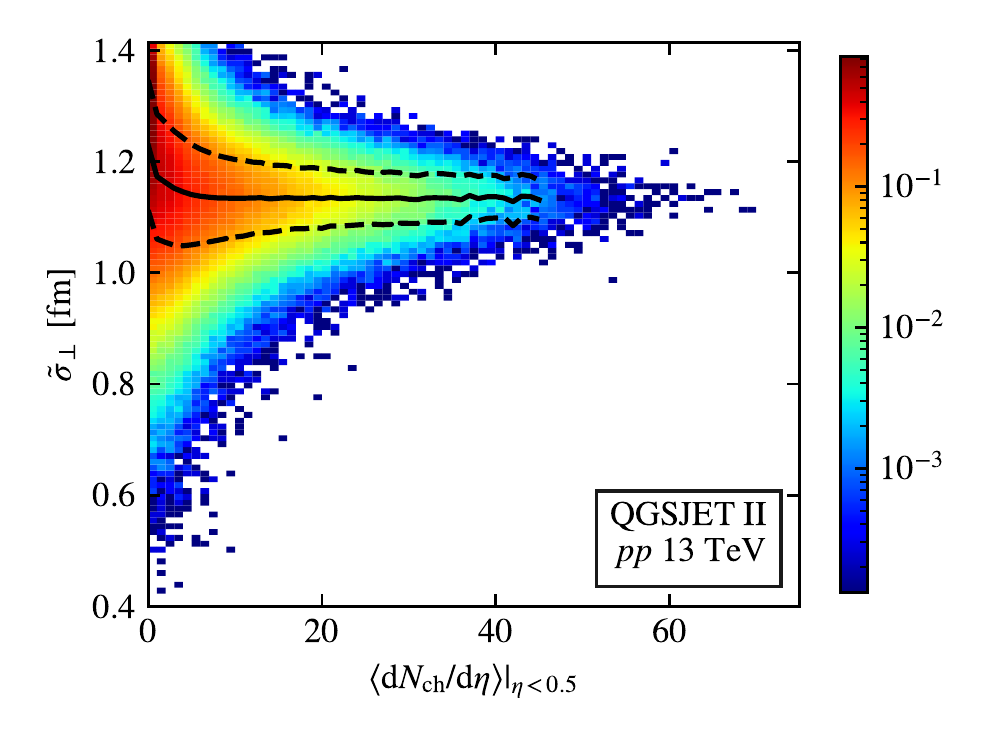}
    \includegraphics[width=\columnwidth]{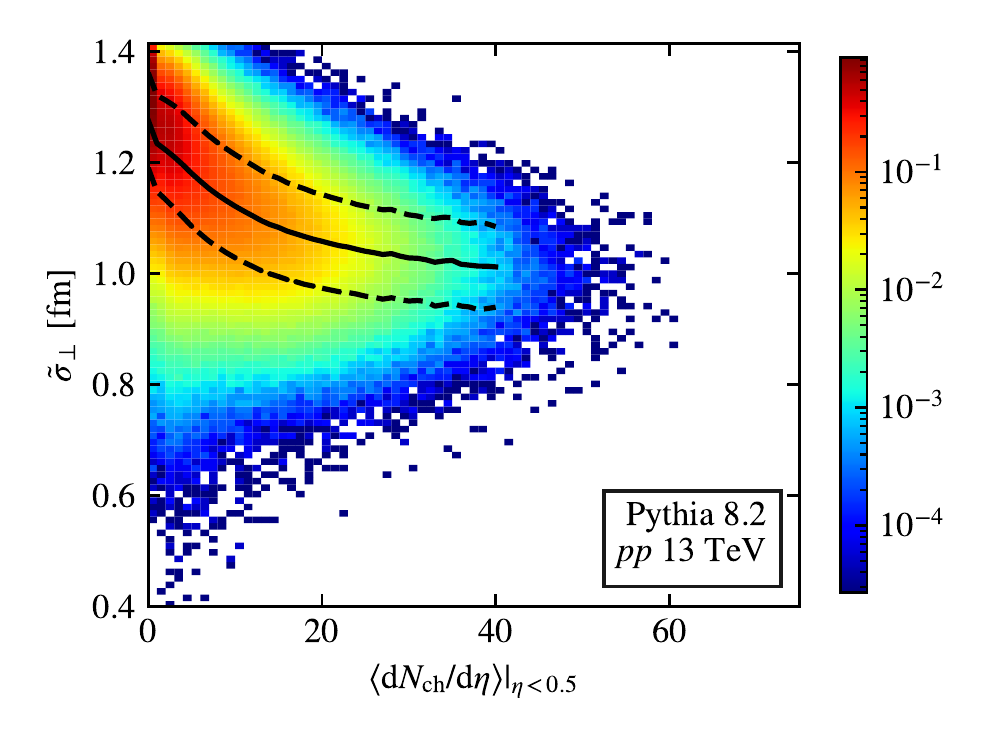}
    \caption{Spread of $\tilde \sigma_\perp$ as a function of the number
    of charged particles in the central pseudo-rapidity region, for
    QGSJET II (above) and Pythia 8.2 (below).
    The mean value of $\tilde \sigma_\perp$ at each $N_\mathrm{ch}$ and
    its standard deviation are shown in solid and dashed lines, respectively;
    the colour code shows the probability density of events with a given
    $\tilde \sigma_\perp$ and $\mathrm{d}N_\mathrm{ch} /\mathrm{d}\eta\;(\eta<0.5)$.}
    \label{fig:pp_sigma}
\end{figure}

In the WiFunC model, the multiplicity dependence emerges due to two-nucleon
momentum correlations and the dependence 
of the emission region of nucleons on the event kinematics. 
In combination, these effects lead to the non-trivial multiplicity
dependence visible in Fig.~\ref{fig:B2}: For increasing multiplicity, the 
momentum phase
space available for single nucleons will on average decrease, which implies an
increased coalescence probability according to Eq.~\eqref{eq:WiFunC}. 
The main multiplicity dependence of the emission region in $pp$ collisions
comes from the  modification  of the transverse 
spread by the  Lorentz boost, as it can be seen from 
Eq.~\eqref{eq:zeta}. In order to get a sense of this 
dependence, we plot in Fig.~\ref{fig:pp_sigma} the multiplicity dependence
of the transverse spread using Pythia and QGSJET at 13\,TeV. In both cases,
$\sigma_\perp=1$\,fm is used. Both event generators lead qualitatively to the
same  multiplicity dependence: The average transverse
momentum increases with increasing number of produced particles, leading to
a decrease in the transverse spread. Such an increase of the
average $p_{T}$ with multiplicity has been observed by all experiments 
at LHC, being reasonably
reproduced by  Pythia (see, e.g., Ref.~\cite{Sjostrand:2020gyg}) and
leading to 
a gradual decrease of $\tilde \sigma_\perp$ up to the rather high values
of $\mathrm{d}N_\mathrm{ch} /\mathrm{d}\eta$. On the other hand, this effect
is not properly described by QGSJET-II, in which case the decrease of 
 $\tilde \sigma_\perp$ is saturated already for relatively small values of
  $\mathrm{d}N_\mathrm{ch} /\mathrm{d}\eta$.

\section{Astrophysical applications}
\label{sec:astrophysical_applications}

Thus far, we have considered only particles at central (pseudo-) rapidity,
which are accessible experimentally. The bulk of produced particles will,
however, in general have large longitudinal momenta. In high energy collisions
at the LHC, the use of a constant $B_2$ as function of $p_z$ is a good
approximation. Therefore, one may naively expect this assumption to be a good
approximation for astrophysical processes as well. This is, as we will discuss
in this section, however not the case. Even so, an isotropic model with
constant $B_A(p_z)$ is still regularly applied in the literature to antinuclei
production in proton-proton 
collisions~\cite{Blum:2017qnn,Korsmeier:2017xzj,Poulin:2018wzu}. 

\begin{figure}[htbp]
    \centering
    \includegraphics[width=0.95\columnwidth]{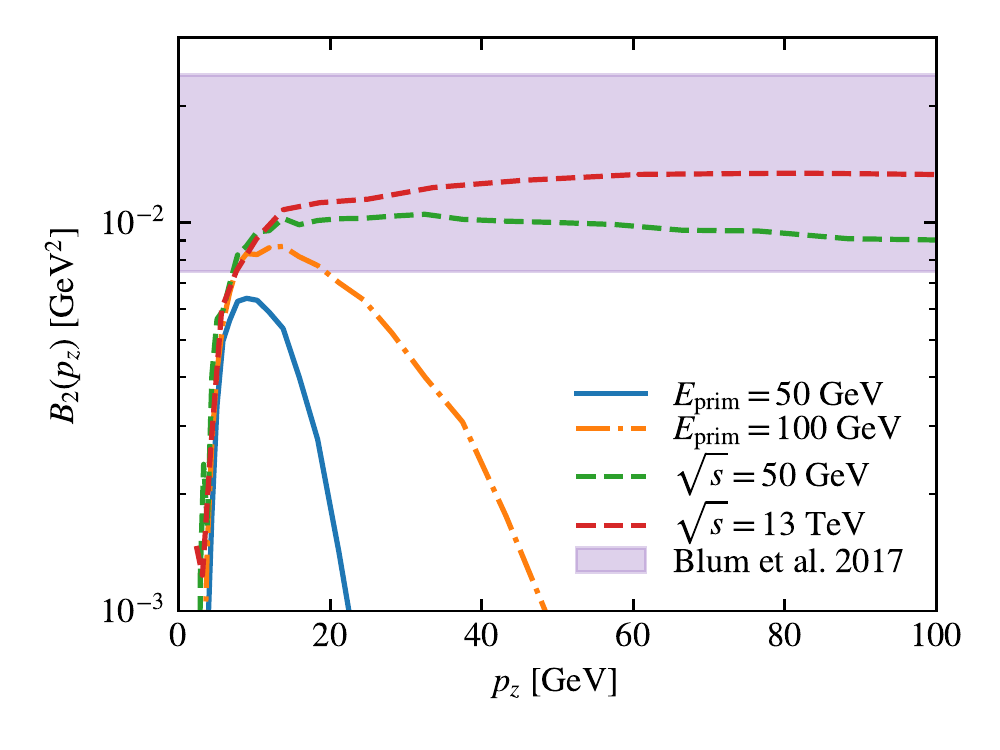}
    \caption{The coalescence factor  for $\bar d$ production,
     as a function of longitudinal momentum
    in the lab frame in $pp$ collisions for various energies relevant
    for astrophysical processes and collider energies.}
    \label{fig:B2_astrophysical}
\end{figure}

Cosmic ray antideuterons are expected to originate in secondary production,
i.e.\ in collisions between primary cosmic rays and the interstellar matter.
The main contribution comes from protons with energies
$E_\mathrm{prim}\sim 20$--$100\unit{GeV}$ colliding with protons in the
interstellar matter, while the bulk of the produced
antideuterons has kinetic energies per nucleon in the range
$T\sim2$--$20\unit{GeV}/n$ \cite{Kachelriess:2020uoh}. In order to check the
validity of a constant $B_2(p_z)$ for astrophysical applications, we therefore
plot the coalescence parameter $B_2(p_z)$ obtained using QGSJET II
for primary energies\footnote{
    Notice that we consider $p_z$,
    $E_\mathrm{prim}$ and $T$ in the rest frame of the target.}
$E_\mathrm{prim} = 50$ and 100\,GeV in Fig.~\ref{fig:B2_astrophysical}
as function of the momentum $p_z$ in the lab frame.
The range of $B_2$  determined using the femtoscopy framework in
Ref.~\cite{Blum:2017qnn} is shown as a violet band.
For comparison, we also show the coalescence factors $B_2$
obtained for $\sqrt{s}=50$\,GeV and 13\,TeV.
In the case of collider energies, the values obtained agree well with the
value inferred by femtoscopy experiments in Ref.~\cite{Blum:2017qnn}. At
energies most relevant for astrophysical processes, however,  the femtoscopy
data at the LHC overestimate the coalescence parameter.
More importantly, the coalescence parameter depends strongly on the
longitudinal momentum at these energies\footnote{The decrease of $B_2$ with
$p_z$ arises mostly from a reduction of the kinematic space available for a 
production of an antinucleon pair. In particular,  $B_2\rightarrow 0$ when
$p_z$ approaches $E_\mathrm{prim}/2$.}. In order to obtain the correct
energy spectra of the produced antinuclei  in astrophysical
processes, a careful treatment taking into
account two-particle correlations is therefore required.

\section{Conclusions}

The WiFunC model is a per-event coalescence model based on the Wigner function
representation of the produced nuclei states, which allows one to account for
both two-nucleon momentum correlations and the size of the hadronic emission
volume. We have shown that this model reproduces well the source size for
baryon emission and the coalescence factor $B_2$ measured recently by the
ALICE collaboration in $pp$ collisions. While these measurements have
characteristics that are often attributed to the collective flow of the
Quark-Gluon Plasma, our results show that the same properties are
well reproduced describing the underlying physical processes  by
conventional QCD inspired event generators as QGSJET or Pythia.
Finally, we have demonstrated that the coalescence parameter depends
strongly on the longitudinal momentum for the  energy range most relevant
for astrophysical processes. Therefore, the use of a constant $B_A$ value
in astrophysical applications should be abandoned.

\begin{acknowledgements}
We are grateful to David Dobrigkeit Chinellato for supplying us with the data
from Ref.~\cite{Acharya:2020sfy} before its official publication.
We thank Bruce Yabsley for pointing out the possible application of the
WiFunC model to exotic bound-states like the $X(3872)$.
\end{acknowledgements}


\end{document}